\documentclass[reprint,amsmath,amssymb,amsart,aps,superscriptaddress,twocolumn,float]{revtex4-2}
\usepackage{graphicx,graphics,float}
\usepackage{dcolumn}
\usepackage{bm}
\usepackage{amssymb}
\usepackage{amsfonts}
\usepackage{float}
\usepackage{xcolor}
\usepackage{hyperref}

\hypersetup{
	pdfnewwindow=true,      
	colorlinks=true,       
	linkcolor=blue,          
	citecolor=blue,        
	filecolor=blue,      
	urlcolor=blue,          
}

\begin{document}

\title{Signature of Magnon-Raman Phonon-Polariton Condensation in a Cavity via Transverse Pumping}
 \author{Subrata Chakraborty}
\email[Correspondence to: ]{subrata@dubai.bits-pilani.ac.in}
\affiliation{Department of General Sciences, Birla Institute of Technology and Science, Pilani-Dubai Campus, Dubai International Academic City, Dubai 345055, UAE\looseness=-1}
 \author{Shishir Kumar Pandey}
\affiliation{Department of General Sciences, Birla Institute of Technology and Science, Pilani-Dubai Campus, Dubai International Academic City, Dubai 345055, UAE\looseness=-1}
\affiliation{Department of Physics, Birla Institute of Technology and Science, Pilani, Hyderabad Campus, Jawahar Nagar, Kapra Mandal, Medchal District, Telangana 500078, India}
 
\date{\today}

\begin{abstract} 
Polaritons---hybrid light-matter quasiparticles---provide a versatile platform for dynamically controlling a wide range of condensed matter systems. While conventional polaritonic platforms rely on direct dipole coupling, controlling dipole-forbidden or Raman-active lattice and spin excitations remains challenging due to optical selection rules and the limitations of THz cavities. Here, we propose a theoretical platform for realizing a continuous-wave magnon--Raman phonon--polariton condensate. By embedding a magnetic medium hosting strongly coupled magnon and Raman-active phonon (MRP) modes inside an optical microcavity under continuous-wave transverse laser pumping, we derive the stationary-state phase diagram and demonstrate the emerging signature of distinct MRP condensation phases under critical conditions. Furthermore, since the bare magnon frequency is tunable via an external magnetic field, we show that the transverse pump frequency and the external magnetic field-dependent magnon simultaneously serve as highly flexible control parameters for exploring and controlling macroscopic quantum phenomena at the interface of cavity quantum optics, lattice dynamics, and quantum magnetism.
\end{abstract}

\maketitle

\section{Introduction}
The engineering of hybrid light-matter quasiparticles, known as polaritons, has emerged as a foundation of modern condensed matter physics and quantum optics. Through the strong coupling between a confined electromagnetic field and elementary electronic or collective excitations in solids, polaritons inherit properties from both their light and matter constituents. Among these, exciton-polaritons in semiconductor microcavities have served as a prominent paradigm for studying non-equilibrium Bose-Einstein condensation (BEC) and macroscopic superfluidity \cite{kasprzak2006boson, carusotto2013quantum}. Concurrently, phonon-polaritons, which arise from the direct coupling of photons to optically active lattice vibrations, offer profound opportunities for mid-infrared and terahertz (THz) photonics, as well as the active manipulation of material properties \cite{basov2016polaritons}. The phonons govern critical collective phenomena such as charge-density waves and conventional superconductivity, modifying phonon dynamics via cavity confinement provides a route for engineering correlated electronic phases \cite{sentef2018cavity, schlawin2022cavity}.

In this context, phonon-polaritons arising from the coupling between photons and collective lattice vibrations hold profound promise for infrared photonics and the active manipulation of lattice degrees of freedom. The standard approaches to phonon-polaritonic engineering face fundamental physical limitations. Direct strong coupling typically requires cavities that support high-quality factors in the THz or infrared frequency domains, which present significant nanofabrication challenges \cite{natcom2026, pra2022}. Selection rules can sometimes strictly limit direct light-matter coupling to infrared-active phonon modes that possess a net dipole matrix element \cite{bourzutschky2024raman}. However, in many correlated materials, the key lattice vibrations mediating electronic phase transitions are dipole-forbidden or exclusively Raman-active. To circumvent these limitations, Bourzutschky and others recently proposed a transformative approach inspired by the concept of ``synthetic cavity quantum electrodynamics (QED)'' widely utilized in cold-atom systems \cite{bourzutschky2024raman, dicke1010superradiance, baumann2010dicke, prl24}. By implementing a continuous-wave transverse-pumping Raman scheme, they demonstrated that an optical-frequency cavity mode can be coupled to \textit{any} Raman-active phonon mode via a two-photon transition \cite{bourzutschky2024raman}. This coupling was mediated by virtual electronic excitations (excitons). Crucially, this scheme enables a fully tunable effective photon-phonon coupling that scales with the external pump strength, paving the way for stationary-state Raman phonon-polariton condensation.

Concurrently, a parallel research area has emerged in the study of collective spin-lattice dynamics within low-dimensional magnetic systems, particularly van der Waals (vdW) magnets \cite{burch2018magnetism,nano20, prl20, prl21, prb21}. These materials often exhibit strong and selective spin-lattice coupling which can lead to a hybridization between collective spin waves (magnons) and Raman phonons due to lattice vibrations. As demonstrated recently by Jana et al., magnetoelastic or anisotropic exchange-mediated interactions in materials like $\text{CoPS}_3$ cause the formation of coherent magnon-Raman phonon hybrid states \cite{jana2025strong}. The manipulation of these hybrid states has traditionally relied on static magnetic fields and elastic strains. However, how synthetic cavity QED platforms can be leveraged to achieve macroscopic magnon- Raman phonon- Polariton (MRP) condensation in magnetic materials under stationary-state condition is worth exploring.
\begin{figure}[h]
\begin{center}
\includegraphics[width=8cm, height=6cm]{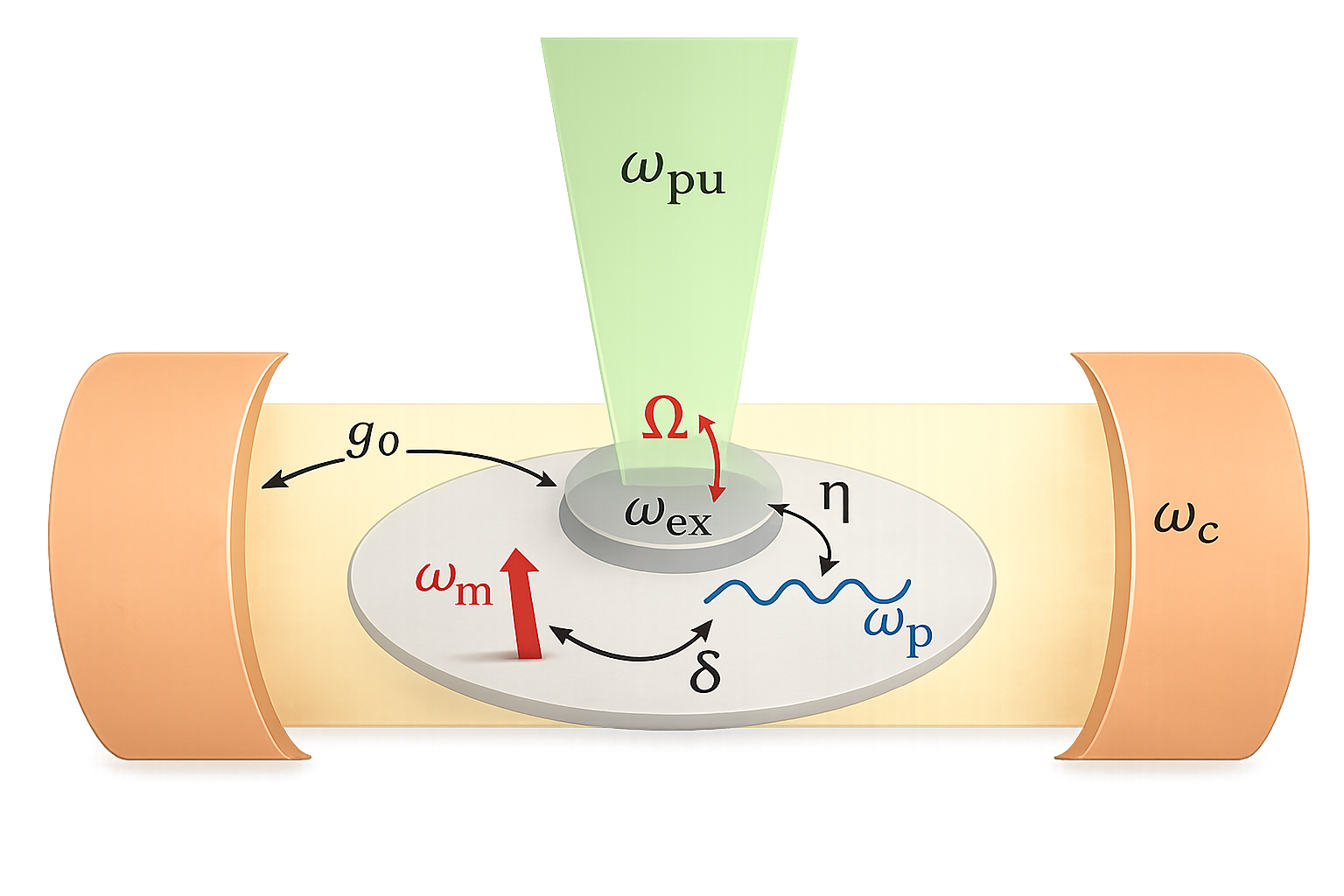}
\caption{\label{schematic}{Schematic of a magnetoelastic material which is placed inside a transversely pumped optical cavity. The cavity mode (light yellow shading), the transverse pump laser (light green shading) and the material medium (gray shading) are shown in the figure. The magnon of frequency $\omega_m$ and the Raman phonon of frequency $\omega_p$ are represented by a thick red arrow and a wavy line, respectively. These two modes are coupled with an interaction strength $\delta$. The excitonic modes of the material with frequency $\omega_{\rm{ex}}$ are represented by the dark gray-shaded region. The exciton--Raman phonon and exciton--cavity coupling strengths are denoted by $\eta$ and $g_0$, respectively. The excitons are further coupled to a transverse pump laser of frequency $\omega_{\mathrm{pu}}$ with a parametric coupling strength $\Omega$.}}
\end{center}
\end{figure} 

In this paper, we bridge these two distinct fields by proposing and analyzing a theoretical platform for realizing a \textit{magnon-Raman phonon-polariton condensate}. We consider a magnetic medium hosting strongly coupled magnon and Raman phonon modes embedded within an optical microcavity and driven by a continuous-wave transverse laser pump. By developing a low-energy effective description that systematically captures the interplay between the electronic excitons, Raman phonons, magnons, and cavity photons, we map out the stationary-state phase diagram of the system. This phase diagram indicates the possible regions where the magnon--Raman phonon--polariton condensation phases emerge. We demonstrate that a finite pumping strength can induce magnon-Raman phonon-polariton condensation in the system. This work establishes a controllable, continuous-wave framework for exploring collective quantum phenomena at the intersection of cavity optics, lattice dynamics, and magnetism.  Furthermore, because the bare magnon frequency is sensitive to external magnetic fields, our proposed setup provides an additional flexible knob to control the condensation in the system. 

The remainder of this manuscript is organized as follows. In Sec.~II, we introduce the full microscopic Hamiltonian and define its associated energy terms. Then we outline the derivation of the stationary-state effective Hamiltonian. In Sec.~III, we present our main results, demonstrating the various possible stationary phases of the proposed model. Finally, we summarize our work in Sec.~IV.

\section{Model description}

\subsection{Hamiltonian}
In this work, we consider a model comprising a cavity photon mode, `$a$', electronic exciton modes, `$X_j$', and a Raman phonon mode, `$p$', as schematically demonstrated in Fig.~\ref{schematic}. The Raman phonon mode is hybridized with a long-wavelength magnon mode, `$m$', in a ferromagnet. Furthermore, we explicitly incorporate the exciton–Raman phonon interaction, while the exciton modes are driven by a continuous-wave transverse pump. Setting $\hbar=1$ we describe the overall system in terms of the following Hamiltonian: 
\begin{eqnarray}
&& \hspace{-2mm} \hat{H}= \hat{H}_{\rm{mp}} + \hat{H}_c +\hat{H}_{\rm{ex}} +\hat{H}_{\rm{p-ex}}
+\hat{H}_{\rm{ex-pum}} +\hat{H}_{\rm{ex-c}}. ~~~~ \label{jun28a1}
\end{eqnarray}
The first terms in Eq.~\eqref{jun28a1} describe the magnon and Raman phonon hybrid system as
\begin{eqnarray}
&& \hat{H}_{\rm{mp}} = \omega_m m^\dag m + \omega_p p^\dag p +\delta(m^\dag p +p^\dag m), \label{jun28a2}
\end{eqnarray}
where  the magnonic ($m^\dagger$/$m$) and the phononic ($p^\dagger$/$p$) modes   are coherently coupled via a magnetoelastic or strain-induced interaction of strength $\delta$ \cite{jana2025strong}. The bare magnon and the phonon modes' energies are $\omega_m$ and $\omega_p$, respectively. We here note that the magnon mode's energy, a flexible parameter, one can control via an external magnetic field. In Eq.~\eqref{jun28a1} the quantized electromagnetic field within the microcavity is modeled by $\hat{H}_c = \omega_c a^\dag a$, at a resonant frequency $\omega_c$. The bare electronic excitations are captured by the third term in Eq.~\eqref{jun28a1} as $\hat{H}_{\rm{ex}}= \omega_{\rm{ex}} \sum_j X_j^\dag X_j$
with $\omega_{\text{ex}}$ signifying the transition frequency of an exciton localized at a lattice site $j$. In Eq.~\eqref{jun28a1} the interaction between the excitonic and lattice degrees of freedom is modeled by a parametric coupling given by \cite{bourzutschky2024raman},
\begin{equation}
    \hat{H}_{\text{p-ex}} = \eta(p^\dagger + p) \sum_j X_j^\dagger X_j, \label{July01a2}
\end{equation}
wherein the phonon displacement field dynamically modulates the local excitonic potential with a coupling strength $\eta$. Semiclassical optical driving of the material medium (excitons) is incorporated through the time-dependent pump term
\begin{equation}
    \hat{H}_{\text{ex-pum}} = \Omega \cos(\omega_{\text{pu}}t) \sum_j \left(X_j^\dagger + X_j\right), \label{July01a3}
\end{equation}
parameterized by the driving frequency $\omega_{\text{pu}}$ and the associated Rabi frequency $\Omega$ \cite{bourzutschky2024raman}. Finally in Eq.~\eqref{jun28a1}, the exciton-cavity field coupling is accounted by the following Hamiltonian
\begin{equation}
    \hat{H}_{\text{ex-c}} = g_0 \sum_j \left(X_j^\dagger + X_j\right)(a^\dagger + a), \label{July01a4}
\end{equation}
where $g_0$ represents mean-field exciton-cavity coupling strength \cite{bourzutschky2024raman}.

\subsection{The Bogolyubov-Tyablikov and rotating frame transformations}
In what follows we now diagonalize the magnon-phonon hybrid Hamiltonian in Eq.~\eqref{jun28a2} via the two-mode Bogolyubov-Tyablikov transformation as
$\hat{H}_{\rm{mp}} = \lambda_+ \psi^\dag_+\psi_+ + \lambda_- \psi^\dag_-\psi_-$, where the eigen energy corresponding to the eigenmode $\psi_\pm$ is
\begin{eqnarray}
&& \lambda_\pm =\frac{1}{2} \left[\left(\omega_m +\omega_p\right) \pm \sqrt{\left(\omega_m-\omega_p\right)^2+4\delta^2}\right], \label{jul02a4}
\end{eqnarray}
and the bare magnon (phonon) operator $m$ ($p$) is transformed in terms of the operators $\psi_\pm$ as follows
\begin{eqnarray}
&& m = \sum_{l=\pm} \frac{\delta}{\sqrt{\left(\lambda_l -\omega_m\right)^2 +\delta^2}} ~ \psi_l,  \label{jul02a2} \\
&& p = \sum_{l=\pm} \frac{\lambda_l -\omega_m}{\sqrt{\left(\lambda_l -\omega_m\right)^2 +\delta^2}} ~\psi_l.\label{jul02a3}
\end{eqnarray}
Therefore, using the Bogolyubov-Tyablikov transformation as in Eqs.~\eqref{jul02a2} and \eqref{jul02a3} we can further write Eq.~\eqref{July01a2}, i.e., the interaction between the exciton and lattice degrees of freedom, as  
\begin{eqnarray}
\hat{H}_{\rm{p-ex}} = \eta \sum_j  \sum_{l=\pm}   \frac{\lambda_l -\omega_m}{\sqrt{\left(\lambda_l -\omega_m\right)^2 +\delta^2}} \left(\psi^\dag_l +\psi_l\right)
X_j^\dag X_j. ~~ \label{jul02a5}
\end{eqnarray}
We here notice that Eq.~\eqref{jul02a5} introduces communication channels between the exciton and the hybrid magnon- Raman phonon modes.

Next, We now transform our total model Hamiltonian, introduced in Eq.~\eqref{jun28a1}, into a rotating frame at the pump frequency $\omega_{\rm{pu}}$, such that $\hat{H}_{\rm{rot}}= U_{\rm{rot}} \hat{H} U_{\rm{rot}}^\dag$, where the rotational operator 
\begin{eqnarray}
U_{\rm{rot}} &=& \exp\left[i\omega_{\rm{pu}}t\left(a^\dag a +\sum_j X_j^\dag X_j\right)\right]. \label{jul02a6}
\end{eqnarray}
Considering on stationary state condition, we now disregard time-dependent components and obtain a simplified expression as
\begin{eqnarray}
&& \hat{H}_{\rm{rot}} = \lambda_+ \psi^\dag_+\psi_+ + \lambda_- \psi^\dag_-\psi_- +\Delta_c a^\dag a 
+ \Delta_{\rm{ex}} \sum_j X_j^\dag X_j \nonumber \\
&& ~~~~~~~ + \sum_j X_j^\dag X_j \left[ \eta_+ \left(\psi^\dag_+ +\psi_+\right) 
+ \eta_- \left(\psi^\dag_- +\psi_-\right) 
\right] \nonumber \\
&& ~~~~~~ + \frac{\Omega}{2}\sum_j \left(X_j +X_j^\dag\right) +g_0\sum_j \left(aX_j^\dag + a^\dag X_j\right), ~~~~~~
\label{jul02a7}
\end{eqnarray}
where the energy coefficient
\begin{eqnarray}
\eta_\pm = \eta\left(\frac{\lambda_\pm -\omega_m}{\sqrt{\left(\lambda_\pm -\omega_m\right)^2 +\delta^2}} \right), \label{jul02a8}
\end{eqnarray}
the exciton detuning energy $\Delta_{\rm{ex}} =\omega_{\rm{ex}} -\omega_{\rm{pu}}$ and the cavity detuning $\Delta_{c} =\omega_{c} -\omega_{\rm{pu}}$.
\begin{figure*}[ht!]
\centering
\includegraphics[width=16.0cm, height=13.0cm]
{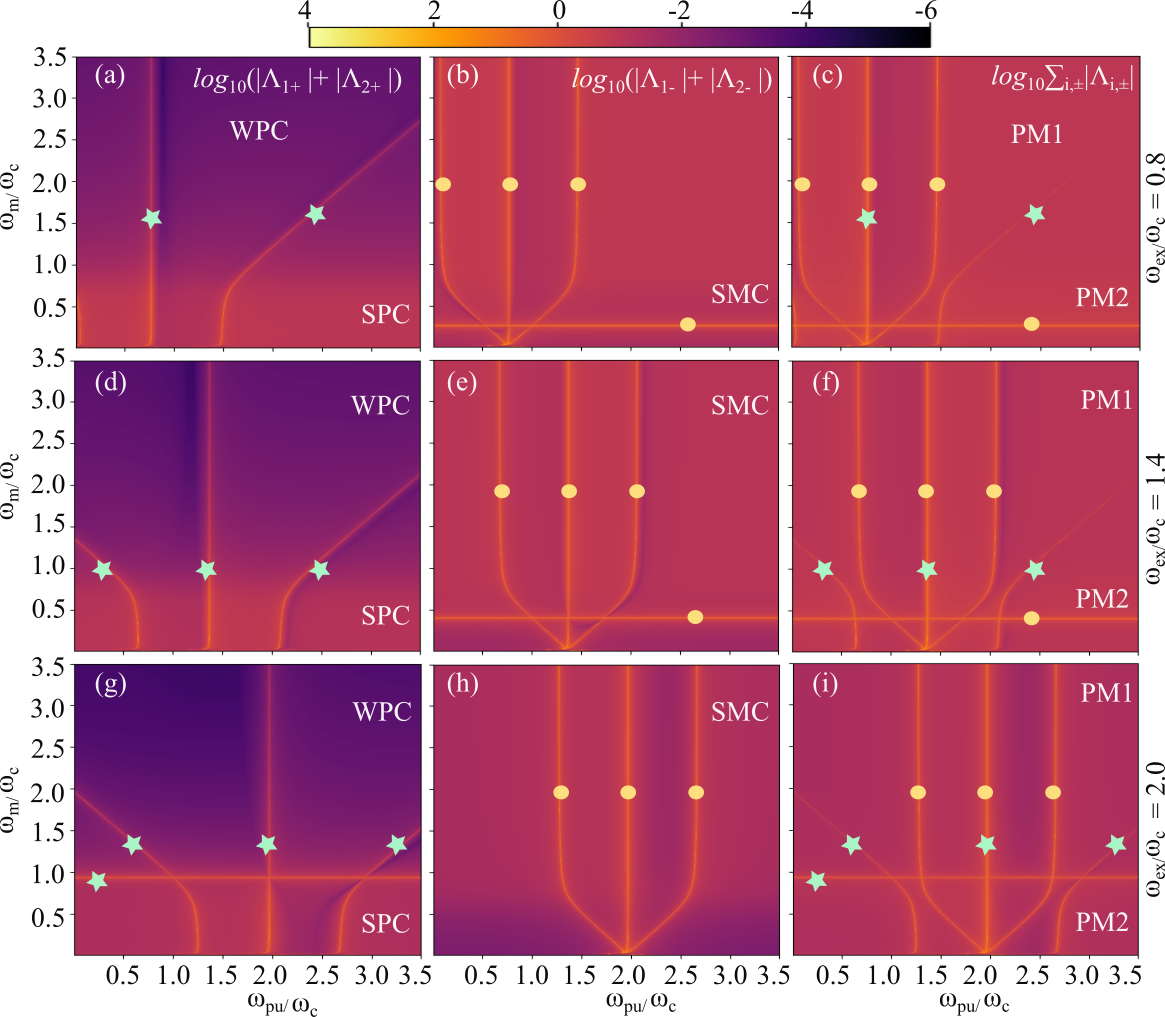}
\caption{Plots of the effective coupling strengths $\Lambda_{1/2\pm}$. The first column [(a), (d) and (g)] and second column [(b), (e) and (h)] show ${\rm{log}}_{10}(|\Lambda_{1+}|+|\Lambda_{2+}|)$ and ${\rm{log}}_{10}(|\Lambda_{1-}|+|\Lambda_{2-}|)$, respectively, while the third column [(c), (f) and (i)] presents the total coupling strength, ${\rm{log}}_{10}\left(\sum_{i=1,2}\sum_{j=\pm}|\Lambda_{ij}|\right)$. The three rows correspond to exciton frequencies $\omega_{\rm ex}/\omega_c=0.8$ (top), $1.4$ (middle), and $2.0$ (bottom). In all panels, the remaining parameters are fixed at $\omega_\text{p}/\omega_\text{c}=0.7$, $\delta/\omega_\text{c}=0.1$, $\eta/\omega_\text{c}=0.15$, $g_0/\omega_\text{c}=0.2$, and $\Omega/\omega_\text{c}=0.5$.
}
\label{fig01}
\end{figure*}

\subsection{The Lang-Firsov polaron transformation}
After the rotating frame transformation we then apply the Lang-Firsov (LF) polaron transformation for the following bosonic operators as follows \cite{jetp63} 
\begin{eqnarray}
&& \hspace{-1.6cm} \bar{\psi}_\pm = e^{-S_{\rm{LF}}} \psi_\pm e^{S_{\rm{LF}}} = \psi_\pm +\frac{\eta_\pm}{\lambda_\pm} \sum_j X_j^\dag X_j,  \label{jul02a9} \\
&& \hspace{-1.6cm} \bar{X}_j = e^{-S_{\rm{LF}}} X_j e^{S_{\rm{LF}}} = X_j \exp \left[\sum_{l=\pm} \frac{\eta_l}{\lambda_l} \left(\psi^\dag_l -\psi_l\right) 
\right],  \label{jul02a10} \\
\rm{with~}&& S_{\rm{LF}} = \sum_j X_j^\dag X_j \left[\sum_{l=\pm} \frac{\eta_l}{\lambda_l} \left(\psi^\dag_l -\psi_l\right)  
\right].  \label{jul02a11}
\end{eqnarray}
Using the LF polaron transformation as in Eqs.~\eqref{jul02a9} and \eqref{jul02a10} we can now express the stationary-state Hamiltonian in the rotating frame as follows $\hat{H}_{\rm{rot}} =\hat{H}_0 +\hat{H}_1$ with
\begin{eqnarray}
 \hat{H}_0 &=& \sum_{l=\pm} \lambda_l {\bar{\psi}}^\dag_l{\bar{\psi}}_l 
  + {\tilde{\Delta}}_{\rm{ex}} \sum_j {\bar{X}}_j^\dag {\bar{X}}_j +\Delta_c a^\dag a,  \label{jul02a12a} \\
 \hat{H}_1 &=&  \left(\frac{\Omega}{2} +g_0~a\right) \sum_j {\bar{X}}_j^\dag  \left[ 1 + \sum_{l=\pm} \delta_l \left(\bar\psi^\dag_l -\bar\psi_l\right)  
\right] \nonumber \\
&& ~~~~~ + ~ {\rm{h.c.}} ~ +O(\delta_+^2,\delta_-^2), ~~ \label{jul02a12}
\end{eqnarray}
where $\delta_{\pm}=\eta_\pm/\lambda_\pm$, ${\tilde{\Delta}}_{\rm{ex}}=\Delta_{\rm{ex}}-\lambda_+\delta_+^2 -\lambda_- \delta_-^2$ is the dressed exciton energy, and h.c. denotes the Hermitian conjugate. 

\subsection{The Schriefer-Wolf transformation}
Following earlier works \cite{bourzutschky2024raman, pra12, pr66}, we next use the Schriefer-Wolf (SW) transformation to describe the system in terms of the effective interactions between the cavity mode `$a$' and the hybrid magnon-phonon `$\bar\psi_\pm$' mode; see Eq.~\eqref{jul02a9} for reference. Considering Eqs.~\eqref{jul02a12a} and \eqref{jul02a12}, the SW transformation needs an appropriate generator $S_{\rm{SW}}$, such that $\left[S_{\rm{SW}} , \hat{H}_0 \right] = -\hat{H}_1$ \cite{bourzutschky2024raman, pra12, pr66}. This transformation converts the rotating frame Hamiltonian $\hat{H}_{\rm{rot}}$ as 
\begin{eqnarray}
 \hat{H}_{\rm{SW}} &=& \exp\left[{S_{\rm{SW}}}\right] \left(\hat{H}_0 +\hat{H}_1 \right)
\exp\left[-{S_{\rm{SW}}}\right] \nonumber \\
&=& \hat{H}_0 +\frac{1}{2} \left[{S_{\rm{SW}} }, \hat{H}_1 \right]. \label{jul02a13}
\end{eqnarray}
Since our primary interest in the present work is to investigate the polariton physics arising from the interactions between the cavity mode `$a$' and the hybrid magnon–phonon modes `$\bar\psi_\pm$', the SW transformation yields the following simplified effective Hamiltonian 
\begin{eqnarray}
 \hat{H}_{\rm{SW}}^{\rm{eff}} &=& \lambda_+ {\bar{\psi}}^\dag_+{\bar{\psi}}_+ + \lambda_-{\bar{\psi}}^\dag_-{\bar{\psi}}_- +E_c a^\dag a \nonumber \\
&& \hspace{-5mm}  + \omega_c \left[ \Lambda_{1+}\left(a \bar\psi_+^\dag + a^\dag \bar\psi_+\right)
+  \Lambda_{2+} \left(a \bar\psi_+ + a^\dag \bar\psi_+^\dag\right) \right. \nonumber \\
&& \hspace{-5mm}  \left. +  \Lambda_{1-}\left(a \bar\psi_-^\dag + a^\dag \bar\psi_-\right)
+  \Lambda_{2-} \left(a \bar\psi_- + a^\dag \bar\psi_-^\dag\right)\right], ~~~~~~~ 
 \label{jul02a14}
\end{eqnarray}
where the renormalized dressed cavity detuning becomes $E_c = \Delta_c - 0.5\sum_j g_0 A_1$ and the coupling strengths in Eq.~\eqref{jul02a14} we find as
\begin{eqnarray}
&& \Lambda_{1\pm} =\frac{g_0\Omega\delta_\pm}{2\omega_c} \left(A_1 -A_{2\pm} -\tilde{A}_1 -\tilde{A}_{3\pm}\right), \label{jul02a15} \\
&& \Lambda_{2\pm} =\frac{g_0\Omega\delta_\pm}{2\omega_c} \left(-A_1 -A_{3\pm} +\tilde{A}_1 -\tilde{A}_{2\pm}\right), \label{jul02a16} 
\end{eqnarray}
where the coefficients are
\begin{eqnarray}
&& \hspace{-3mm} A_1 = \frac{1}{\tilde{\Delta}_{\rm{ex}}-\Delta_c}, ~ {\tilde{A}}_1 = \frac{1}{2\tilde{\Delta}_{\rm{ex}}},
  \nonumber \\
&& \hspace{-3mm} A_{2\pm} = \frac{1}{\tilde{\Delta}_{\rm{ex}}+\lambda_\pm -\Delta_c}, {\tilde{A}}_{2\pm} = \frac{1}{2\left(\tilde{\Delta}_{\rm{ex}}+\lambda_\pm \right) }, \nonumber \\
&& \hspace{-3mm} A_{3\pm} = \frac{1}{-\tilde{\Delta}_{\rm{ex}}+\lambda_\pm +\Delta_c}, {\tilde{A}}_{3\pm} = \frac{1}{2\left(-\tilde{\Delta}_{\rm{ex}}+\lambda_{\pm} \right)}. \label{jul20a1} ~~~~~
\end{eqnarray}
Equation~\eqref{jul02a14} suggests that if at least one of the two coupling strengths, $\Lambda_{1+}$ or $\Lambda_{2+}$, becomes singular, the system can exhibit condensation between the magnon--Raman phonon hybrid mode `$\bar{\psi}_{+}$' and the cavity mode `$a$'. From Eqs.~\eqref{jul02a15}-\eqref{jul20a1}, it follows that this occurs when at least one of the denominators of $A_{1}$, $\tilde{A}_{1}$, $A_{2/3+}$, or $\tilde{A}_{2/3+}$ approaches zero. Similarly, if at least one of the two coupling strengths, $\Lambda_{1-}$ or $\Lambda_{2-}$, becomes singular, condensation can occur between the magnon--Raman phonon hybrid mode `$\bar{\psi}_{-}$' and the cavity mode `$a$'. This happens when at least one of the denominators of $A_{1}$, $\tilde{A}_{1}$, $A_{2/3-}$, or $\tilde{A}_{2/3-}$ vanishes. In addition, we further note that if at least one of the coefficients $A_{1}$ or $\tilde{A}_{1}$ becomes singular, the system can exhibit simultaneous condensations between the `$\bar{\psi}_{+}$' and `$a$' modes, and between the `$\bar{\psi}_{-}$' and `$a$' modes.
 
\section{Results and discussion}
In what follows, we demonstrate various stationary phases of the magnon-phonon hybrid system inside a transversely pumped cavity. These phases are presented in Figs.~\ref{fig01}(a)–\ref{fig01}(i) as functions of the externally controllable parameters $\omega_m$ and $\omega_{\rm{pu}}$, with the model parameters set to $\omega_\text{p}/\omega_\text{c} = 0.7$, $\delta/\omega_\text{c} = 0.1$, $\eta/\omega_\text{c} = 0.15$, $g_0/\omega_\text{c} = 0.2$, and $\Omega/\omega_\text{c} = 0.5$.

Figures~\ref{fig01}(a), \ref{fig01}(d), and \ref{fig01}(g) demonstrate $\log_{10}\!\left(\left|\Lambda_{1+}\right|+\left|\Lambda_{2+}\right|\right)$ for three different excitation frequencies, $\omega_{\mathrm{ex}}$. The bright lines (marked by green stars) indicate the condensation of the hybrid magnon--Raman phonon mode `$\bar{\psi}_{+}$' together with the cavity mode `$a$'. This magnon--phonon--polariton condensation appears here via giant (singular) effective coupling $\Lambda_{1/2+}$ between the hybrid `$\bar{\psi}_{+}$' mode and the cavity mode `$a$'. Furthermore, in Figs.~\ref{fig01}(a), \ref{fig01}(d), and \ref{fig01}(g), we also identified two distinct coupling regimes: a weakly hybridized `$\bar{\psi}_{+}$' and `$a$' phase (WPC), and a stronger hybridized `$\bar{\psi}_{+}$' and `$a$' phase (SPC). The WPC phase appears at relatively higher magnon energies, where effective weak magnon--phonon hybridization produces negligible polariton formation. On the other hand, the SPC phase is characterized by stronger magnon--phonon--polariton hybridization, arising from enhanced exciton-mediated interactions among the magnon, phonon, and cavity modes.

In Figs.~\ref{fig01}(b), \ref{fig01}(e), and \ref{fig01}(h), we plot ${\rm{log}}_{10}\left(|\Lambda_{1-}|+|\Lambda_{2-}|\right)$ as a function of $\omega_m$ and $\omega_{\rm{pu}}$. The bright lines (highlighted by yellow circles) indicate the condensation of the hybrid magnon--Raman phonon mode `$\bar{\psi}_{-}$' together with the cavity mode `$a$'. This magnon--phonon--polariton condensation arises from the giant (singular) effective couplings $\Lambda_{1/2-}$ between the hybrid `$\bar{\psi}_{-}$' mode and the cavity mode `$a$'. Furthermore, Figs. \ref{fig01}(b), \ref{fig01}(e), and \ref{fig01}(h) also identify the SMC phase, characterized by considerably stronger hybridization between the hybrid magnon--phonon mode `$\bar{\psi}_{-}$' and the cavity mode `$a$'. The SMC phase emerges due to sizable effective coupling between the hybrid magnon--phonon mode `$\bar{\psi}_{-}$' and the cavity mode `$a$', mediated by the exciton-induced phonon--cavity interaction and resulting a magnon- phonon- polariton formation. We note that when the magnon energy deviates significantly from the phonon energy, the magnon and phonon become effectively decoupled, and the SMC predominantly corresponds to phonon--polariton formation.

Finally, by combining Figs. \ref{fig01}(a)–\ref{fig01}(b), \ref{fig01}(d)–\ref{fig01}(e), and \ref{fig01}(g)–\ref{fig01}(h), we obtain Figs. \ref{fig01}(c), \ref{fig01}(f), and \ref{fig01}(i), respectively (i.e., the rightmost column of Fig. \ref{fig01}). In these figures, the bright lines represent two distinct condensates: the $\bar{\psi}_{+}$ and $a$ condensate (indicated by green stars); and the $\bar{\psi}_{-}$ and $a$ condensate (indicated by yellow circles). In the rightmost column of Fig. \ref{fig01}, we note that some of the bright lines simultaneously indicate the signatures of the both condensates. Additionally, in the rightmost column of Figs. \ref{fig01}, we also indicate PM1 and PM2 phases correspond to the coexistence of the WPC and SMC phases, and the SPC and SMC phases, respectively.

\section{Conclusion}
In conclusion, we bridge cavity quantum optics, lattice dynamics, and magnetism by proposing a theoretical platform for a magnon-Raman phonon-polariton condensate. The system consists of a magnetic medium that hosts strongly coupled magnon and Raman-active phonon modes, placed inside an optical microcavity and driven by a continuous-wave transverse laser pump.
A low-energy effective Hamiltonian is developed that systematically incorporates the interplay among electronic excitons, Raman phonons, magnons, and cavity photons. This framework allows us to derive the stationary-state phase diagram of a magnon-phonon hybrid system inside a cavity via transverse pumping. We demonstrate the signatures of various stationary-state magnon–Raman phonon–polariton condensations under certain critical conditions. We show that, in addition to the pump laser frequency, the strong dependence of the bare magnon frequency on the external magnetic field provides an additional highly flexible control parameter. These tunable parameters enable the precise exploration and manipulation of the condensation transition. The proposed architecture establishes a fully continuous-wave, magnetically controllable platform for investigating collective quantum phenomena at the intersection of light, spin, and lattice degrees of freedom.

S.C. and S.K.P. acknowledge the NFSG grant from BITS-Pilani, Dubai campus, which supported this research.


%

\end{document}